\documentclass[a4paper,superscriptaddress,prl,twocolumn]{revtex4}
\RequirePackage{graphicx}
\renewcommand{\d}{\mathrm d}
\begin{document}
\title{Weak Turbulent Kolmogorov Spectrum for Surface Gravity Waves}

\date{\today}

\author{A.~I.~Dyachenko}
\email{alexd@landau.ac.ru}
\affiliation{Landau Institute for Theoretical Physics,\\
2, Kosygin Str., Moscow, 119334, Russian Federation}

\author{A.~O.~Korotkevich}
\email{kao@landau.ac.ru}
\affiliation{Landau Institute for Theoretical Physics,\\
2, Kosygin Str., Moscow, 119334, Russian Federation}

\author{V.~E.~Zakharov}
\affiliation{Landau Institute for Theoretical Physics,\\
2, Kosygin Street, Moscow, 119334, Russian Federation}
\affiliation{University of Arizona,\\
Tucson, AZ 85721, USA}
\affiliation{Waves and Solitons LLC,\\
738 W. Sereno Dr., Gilbert, AZ 85233, USA}

\begin{abstract}
We study the long-time evolution of surface gravity waves on deep water excited by a stochastic
external force concentrated in moderately small wave numbers. We numerically implemented
the primitive Euler equations for the potential flow of an ideal fluid with free surface written
in Hamiltonian canonical variables, using the expansion of the Hamiltonian in powers of nonlinearity of up to
terms of fourth order.

We show that due to nonlinear interaction processes a stationary Fourier-spectrum of a surface elevation close
to $<|\eta_k|^2> \sim k^{-7/2}$ is formed. The observed spectrum can be interpreted as a
weak-turbulent Kolmogorov spectrum for a direct cascade of energy.
\end{abstract}

\maketitle

Kolmogorov was born in 1903. Now, in the year of his centenary, his greatness is obvious
not only for pure and applied mathematicians. All physicists appreciate his pioneering works
on powerlike cascade spectra in turbulence of the incompressible fluid
\cite{Kolmogorov-1941}.
It is obvious now that cascade processes, similar to the Kolmogorov cascade of
energy, play a very important role in many different fields of physics, such as nonlinear
optics~\cite{Dyachenko-1992}, plasma physics~\cite{Galtier-2002}, hydrodynamics of
superfluid $\mathrm{He}^{4}$, and so forth.

In all these cases the physical situations are similar. There is an ensemble of slowly
decaying, weakly nonlinear waves in a medium with dispersion. Such systems have to be
described statistically. However this is not traditional statistical mechanics, because the
ensembles are very far from thermodynamic equilibrium. Nevertheless, one can develop
a systematic approach for the statistical study of weakly nonlinear waves. This is the theory of
weak (or wave) turbulence~\cite{Springer-1992}. The main tools here are the kinetic
equations for squared wave amplitudes. These equations describe the nonlinear resonant
interaction processes taking place in the wave systems. Like in the turbulence in
incompressible fluid, these processes lead to the cascades of some constants of motion
(energy, wave action, momentum etc.) along the $k-$space. In isotropic systems it might
be either a direct cascade of energy from small to large wave numbers or an inverse cascade of
wave action to small wave numbers~\cite{Zakharov-PhD}. In an anisotropic system the
situation could be much more complicated~\cite{Rossby-waves}.

The brilliant conjecture of Kolmogorov still is a hypothesis, supported by ample experimental
evidence. On the contrary, the existence of powerlike Kolmogorov spectra, describing
cascades in weak turbulence, is a rigorous mathematical fact. These spectra are the exact
solutions of the stationary homogeneous kinetic equation, completely different from the
thermodynamic Rayleigh-Jeans solutions.

Nevertheless, the case is not closed. The weak turbulent theory itself is based on some
assumptions, like phase stochasticity and the absence of coherent structures. This is the reason
why justification of weak turbulent theory is an urgent and important problem.

This justification can be done by a direct numerical solution of the primitive dynamic equation
describing the wave ensemble. In pioneering works by Majda, McLaughlin and Tabak~\cite{MMT} it was done for the 1-D wave system. The results obtained by these authors
are not easily interpreted. In some cases they demonstrate Kolmogorov-type spectra,
in other cases --- power spectra with essentially different exponents.

In article~\cite{Dias-2001-1} deviation from weak turbulent theory was explained by the
role of coherent structures (solitons, quasi-solitons and collapses). If a 1-D system is free
from coherent structures, weak-turbulent spectra are observed with a good deal of
evidence~\cite{Vasiliev-2002,Dias-2001-2,Dysthe-2003}.

In spite of their heuristic value, the 1-D models so far developed have no direct physical application. Real
physical systems, where wave turbulence is realized, are at least 2-dimensional. The most
natural and important examples are capillary and gravity waves on deep water.
A weak-turbulent theory of capillary waves was developed by Zakharov and Filonenko in
1967~\cite{Zakharov-1967}, who found that the correlation function of elevation $\eta(\vec r,t)$
has to be $<|\eta_{\vec k}|^2> \sim k^{-19/4}$. This result was supported by laboratory
experiments, performed independently by three groups (in UCLA~\cite{UCLA},
Niels Bohr Institute~\cite{Niels-Bohr} and the Solid State Physics Institute in Chernogolovka,
Russia~\cite{Kolmakov-Lett, Kolmakov}). The spectrum
$k^{-19/4}$ was obtained by a direct numerical simulation of Euler equation for
incompressible fluid with free surface by Pushkarev and Zakharov~\cite{Pushkarev-1996,Pushkarev-2000,Pushkarev-1999B}.

The most interesting example of 2-D wave ensembles demonstrating weak-turbulent
cascades is a system of gravity waves on the surface of deep water. We are sure that a
weak-turbulent theory of these waves is key to understanding the processes in a
wind-driven sea. However, we do not concentrate on this point in our article.

Our initial goal was to reproduce (and emulate), for gravity waves, the work which was done
by Pushkarev and Zakharov~\cite{Pushkarev-1996} for capillary waves. One has to expect
that this is a more difficult problem, because the leading process in capillary waves is a three-wave interaction
(dispersion of waves is of decay type), while for
gravity waves the lowest order process is four-wave interaction (dispersion of waves is of non decay type).

Attempts to perform direct numerical simulations of potential flow in an ideal fluid with a free
surface were made by several authors~\cite{Tanaka-2001}. Only in one article authors did pay
interest to Kolmogorov-type weak-turbulent spectra~\cite{Onorato-2002}. Authors of this
paper observed the formation of Kolmogorov tails during the time evolution of an artificially
cut-off JONSWAP (Joint North Sea Wave Project, described in~\cite{JONSWAP}) energy spectrum. The results of present article
agree with the results of~\cite{Onorato-2002} completely. However we would like to stress
a difference.

In our work, in contrast to~\cite{Onorato-2002}, we study a forced turbulence, excited by external sources, posed in moderately low
wave numbers. We show that the growth of wave energy due to this forcing is arrested by
nonlinear resonant four-wave processes, which leads to the formation of a powerlike Kolmogorov
spectrum in the inertial interval. In this sense our article is a direct numerical
confirmation of the weak-turbulent theory for surface gravity waves.

{\it Theoretical Background --- }
We study the potential flow of an ideal inviscid incompressible fluid with the velocity potential
$\phi=\phi(x,y,z;t)$
$$
\Delta \phi = 0,
$$
in the infinitely deep domain occupied by the fluid. Equations for the boundary conditions
at the surface are the following
\begin{equation}
\label{Laplas_boundary}
\begin{array}{c}
\displaystyle
\left.
\left(
\dot \eta + \phi'_x \eta'_x + \phi'_y \eta'_y\right)\right|_{z= \eta}
= \left. \phi'_z \right|_{z= \eta},\\
\displaystyle
\left. \left ( 
\dot \phi +
\frac{1}{2}\left|\nabla \phi \right|^2
\right ) \right |_{z= \eta} + g \eta = 0.
\end{array}
\end{equation}
Here $\eta(x,y;t)$ is the surface elevation with respect to still water, $g$ is the gravity
acceleration. Equations (\ref{Laplas_boundary}) are Hamiltonian~\cite{Zakharov-1968}
with the canonical variables $\eta(x,y;t)$ and $\psi(x,y;t)=\phi(x,y,\eta(x,y;t);t)$
\begin{equation}
\label{Hamiltonian_equations}
\frac{\partial \eta}{\partial t} = \frac{\delta H}{\delta \psi}, \;\;\;\;
\frac{\partial \psi}{\partial t} = - \frac{\delta H}{\delta \eta},
\end{equation}
where $H$ is the Hamiltonian of the system
$$
H = \frac{1}{2} \int_{-\infty}^{+\infty} \d x \d y \left(
g\eta^2 + 
\int_{-\infty}^{\eta} |\nabla \phi|^2 \d z
\right),
$$
Unfortunately $H$ cannot be written in the close form as a functional of $\eta$ and $\psi$.
However one can limit Hamiltonian by first three terms of powers of $\eta$ and $\psi$
\begin{equation}
\label{Hamiltonian}
\begin{array}{l}
\displaystyle
H = H_0 + H_1 + H_2 + ...,\\
\displaystyle
H_0 = \frac{1}{2}\int\left( g \eta^2 + \psi \hat k  \psi \right) \d x \d y,\\
\displaystyle
H_1 =  \frac{1}{2}\int\eta\left[ |\nabla \psi|^2 - (\hat k \psi)^2 \right] \d x \d y,\\
\displaystyle
H_2 = \frac{1}{2}\int\eta (\hat k \psi) \left[ \hat k (\eta (\hat k \psi)) + \eta\Delta\psi \right] \d x \d y.
\end{array}
\end{equation}
Here $\hat k$ is a linear integral operator 
$\left(\hat k =\sqrt{-\Delta}\right)$, such that in $k$-space it corresponds to
multiplication of Fourier harmonics ($\psi_{\vec k} = \frac{1}{2\pi} \int \psi_{\vec r} e^{i {\vec k} {\vec r}} \d x \d y$)
by $\sqrt{k_{x}^2 + k_{y}^2}$. For gravity waves this
reduced Hamiltonian describes four-wave interaction. Then dynamical equations (\ref{Hamiltonian_equations}) acquire the form
\begin{equation}
\label{eta_psi_system}
\begin{array}{lcl}
\displaystyle
\dot \eta &=& \hat k  \psi - (\nabla (\eta \nabla \psi)) - \hat k  [\eta \hat k  \psi] +\\
\displaystyle
		&&+ \hat k (\eta \hat k  [\eta \hat k  \psi]) + \frac{1}{2} \Delta [\eta^2 \hat k \psi] + 
		\frac{1}{2} \hat k [\eta^2 \Delta\psi],\\
\displaystyle
\dot \psi &=& - g\eta - \frac{1}{2}\left[ (\nabla \psi)^2 - (\hat k \psi)^2 \right] - \\
\displaystyle
		&& - [\hat k  \psi] \hat k  [\eta \hat k  \psi] - [\eta \hat k  \psi]\Delta\psi + D_{\vec r} + F_{\vec r}.
\end{array}
\end{equation}
Here $D_{\vec r}$ is a dissipation term, which consists of hyperviscosity on small scales
and "artificial" damping on large scales.
$F_{\vec r}$ is the driving term which simulates pumping on large scales (for example, due to wind).
In the $k$-space supports of $D_{\vec k}$ and $F_{\vec k}$ are separated by the inertial
interval, where the Kolmogorov-type solution can be recognized.

Another approach based on boundary integral approximation has been studied by Clamond and Grue in~\cite{Grue-2001}.

We study numerically the quasi stationary solution of equations (\ref{eta_psi_system}).
According to the theory of  weak turbulence, the Fourier spectrum of the surface elevation
averaged by ensemble (corresponding to the flux of energy from large scales to small scales) is
\begin{equation}
\label{desired_spectrum}
<|\eta_k|^2> = \frac{C g^{1/2} P^{1/3}}{k^{7/2}}.
\end{equation}
Here $P$ is the energy flux, and $C$ is a dimensionless Kolmogorov constant. This spectrum is a stationary solution of the wave kinetic equation~\cite{Zakharov-1967}.

It is worth to say that an alternative spectrum was proposed earlier by
Phillips~\cite{Phillips-1958}. 
In contrast to weak turbulent theory Phillips power-like spectrum can be formed due to a
wave breaking mechanism and leads to a surface elevation spectrum
$<|\eta_k|^2> \sim k^{-4}$. It can be realized for higher level of turbulence, which is far
beyond the weak turbulent theory.

{\it Numerical Simulation --- }
For numerical integration of (\ref{eta_psi_system}) we used the following pumping and
dissipation terms which are defined in Fourier space as
\begin{equation}
\label{Damp_and_pump}
\begin{array}{l}
\displaystyle
F_k = f_k e^{iR_{\vec k} (t)},\\
f_k = \cases{
4 F_0 \frac{(k-k_{p1})(k_{p2}-k)}{(k_{p2} - k_{p1})^2}, if (k_{p1} < k < k_{p2});\cr
f_k = 0 - otherwise;\cr}\\
\displaystyle
D_{\vec k} = \gamma_k \psi_{\vec k},\\
\displaystyle
\gamma_{k} = \cases{
-\gamma_1, k \le k_{p1},\cr
0, k_{p1} < k < k_{p2},\cr
- \gamma_2 (k - k_d)^2, k > k_d.\cr}
\end{array}
\end{equation}
Here $R_{\vec k} (t)$ is a uniformly distributed random number in the interval $(0,2\pi)$
for each $\vec k$.
We have applied an implicit difference scheme that keeps the main property of this system ---
conservation of Hamiltonian in the absence of pumping and
damping.

The equations (\ref{eta_psi_system}) were numerically simulated in the periodic domain
$2\pi\times2\pi$. The size of the grid was $512\times512$ points. Gravity
acceleration $g$ was equal to one.
Parameters of the damping and pumping in (\ref{Damp_and_pump})
were the following: $k_{p1} = 5,\; k_{p2} = 10,\;
k_d = 100$. Thus the inertial interval is equal to a decade.

In the simulations we paid special attention to the problems which could ''damage'' the
calculations.
First of all, it is the ''bottle neck'' phenomenon which was studied in the paper~\cite{Falkovich}.
The effect consists in accumulation of energy in $k$-space ahead of the dissipation region,
if the dissipation is too large.
This effect is very fast, but can be effectively suppressed by a proper choice of damping value
$\gamma_2$ in dissipation (\ref{Damp_and_pump}) in the case of moderate pumping values $F_0$.
On the other hand the $F_0$ value should not be too small to secure wave cascade on
the discrete grid.
For the case of capillary waves it was examined in detail in~\cite{Capillary-2003}.
The second problem is the accumulation of ''condensate'' in low wave numbers due to inverse cascade.
Buildup of condensate can be overcome by simple
adaptive damping in the small wave numbers.

After some time the system reaches the
stationary state,
where the balance between pumping and damping takes place. In
this state an important parameter is the ratio of nonlinear energy
to the linear one $(H_1 + H_2)/H_0$.

\begin{figure}[!htbp]
\includegraphics[width=8.5cm]{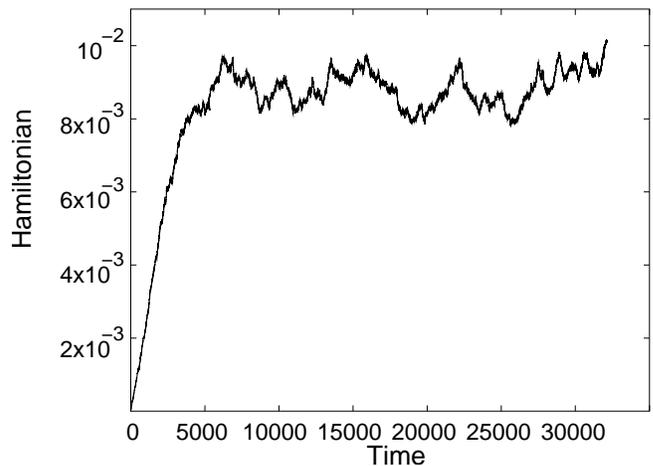}
\caption{\label{Hamiltonian_t}Hamiltonian as a function of time. Quasi stationary state is formed at $t \simeq 5000$.}
\end{figure}

For example, for the external force $F_0 = 2\times10^{-4}, \gamma_1 = 1\times10^{-3},
\gamma_2 = 665$
the level of nonlinearity was equal to
$(H_1 + H_2)/H_0 \simeq 3\times10^{-3}$. The Hamiltonian as a function of time is shown
in Fig.~\ref{Hamiltonian_t}.

\begin{figure}[!htbp]
\includegraphics[width=8.5cm]{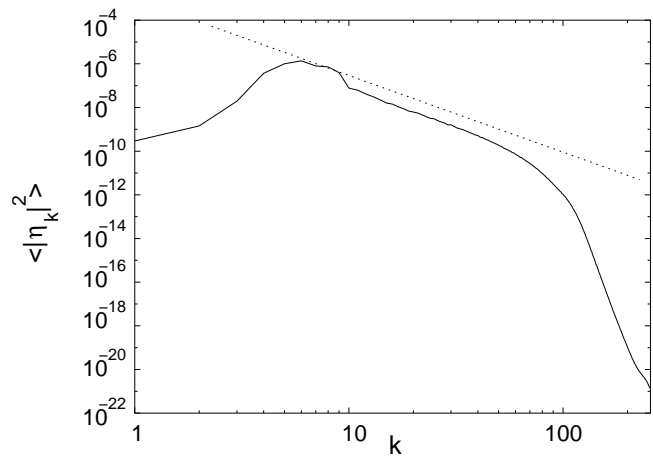}
\caption{\label{etacorr}Averaged spectrum of surface elevation $<|\eta_k|^2>$. Line $\sim k^{7/2}$ is also shown.}
\end{figure}

The averaged spectrum of surface elevation $<|\eta_k|^2>$ appears to be power-like in the essential part of
inertial interval, where the influence of pumping and damping was small. This spectrum is
shown in Fig.~\ref{etacorr}.

\begin{figure}[!htbp]
\includegraphics[width=8.5cm]{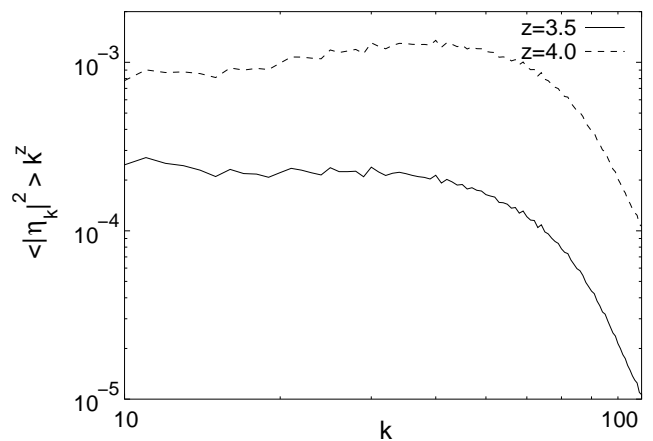}
\caption{\label{compspec}Compensated waves spectra. Solid line --- weak turbulent theory; dashed line --- Phillips theory.
Only one decade in $k$ is shown.}
\end{figure}

One can estimate the exponent of the spectrum. 
Compensated spectra (e.g. multiplied by $k^z$, to get horizontal line)
are shown in Fig.~\ref{compspec}. This figure seems to be the evidence
that the Kolmogorov spectrum predicted by weak turbulence theory fits better the results of the
numerical experiment.

\begin{figure}[!htbp]
\includegraphics[width=8.5cm]{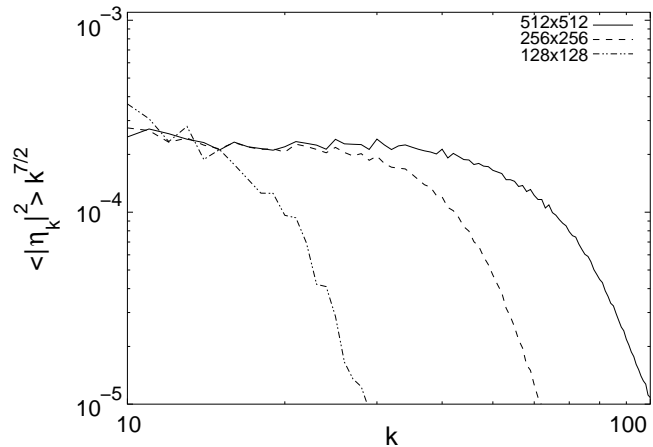}
\caption{\label{broadening}Compensated spectra for the different grids.}
\end{figure}

The quality of the result (closeness to the $<|\eta_k|^2> \sim k^{-7/2}$) crucially depends on
the width of the inertial interval.
In our previous work~\cite{Gravity-2003} similar simulations were performed on the
grid $256\times256$. Weak turbulent spectrum is clearly seen on the grid
$512\times512$, can be hinted at the grid $256\times256$, and is almost invisible on the grid
$128\times128$. This difference is demonstrated in Fig.~\ref{broadening}.

\begin{figure}[!htbp]
\includegraphics[width=8.5cm]{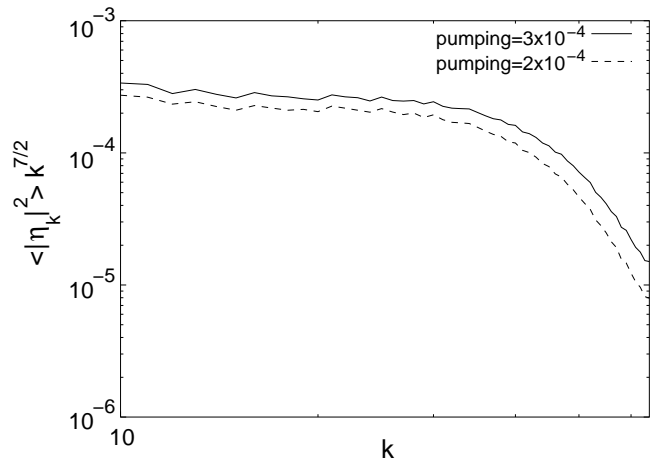}
\caption{\label{diffpump}Compensated spectra for the different pumping level. Solid line --- $F_0 = 3\times10^{-4}$,
dashed line --- $F_0 = 2\times10^{-4}$.}
\end{figure}

In the end we would like to mention that for the different pumping level surface elevation spectra differ only due to the different energy flux value $P$ in (\ref{desired_spectrum}),
as clearly seen in Fig.~\ref{diffpump}.

\begin{acknowledgments}
This work was
supported by ONR grant N00014-03-1-0648, RFBR grant 03-01-00289, INTAS grant 00-292, the Programme
``Nonlinear dynamics and solitons'' from the RAS Presidium and ``Leading Scientific
Schools of Russia" grant, also by US Army Corps of Engineers
Grant DACW 42-03-C-0019 and by NSF Grant NDMS0072803.

Also authors want to thank the creators of the open-source fast Fourier transform library
FFTW~\cite{FFTW} for this fast, portable and completely free piece of software.
\end{acknowledgments}

\end{document}